\documentclass[
 aip,
 jcp,
amsfonts,
amsmath,
amssymb,
reprint,%
]{revtex4-1}

\usepackage[colorlinks,linkcolor=blue,urlcolor=blue,citecolor=blue]{hyperref}
\usepackage{bm}
\usepackage{graphicx}
\usepackage{subfig}
\usepackage{array}

\usepackage{algorithm}
\usepackage{algpseudocode}

\def\bL{\mathbf{L}}
\def\bX{\mathbf{X}}
\def\bY{\mathbf{Y}}
\def\bW{\mathbf{W}}
\def\bM{\mathbf{M}}
\def\bZ{\mathbf{Z}}
\def\bB{\mathbf{B}}
\def\bC{\mathbf{C}}

\def\bVx{\mathbf{V}_{1}}
\def\bVy{\mathbf{V}_{2}}
\def\bVz{\mathbf{V}_{3}}

\def\bk{\mathbf{k}}
\def\bpsi{\bm \Psi}
\def\bxi{\bm \xi}
\def\bchi{\bm \chi}
\def\bPhi{\bm \Phi}

\usepackage{braket}

\newlength{\bracewidth}

\begin{document}

\title{GPU acceleration of hybrid functional calculations in the SPARC electronic structure code}
\author{Xin Jing}
\affiliation{College of Engineering, Georgia Institute of Technology, Atlanta, GA 30332, USA}
\affiliation{College of Computing, Georgia Institute of Technology, Atlanta, GA 30332, USA}
\author{Abhiraj Sharma}
\affiliation{Physics Division, Lawrence Livermore National Laboratory, Livermore, CA, 94550, USA}
\author{John E. Pask}
\affiliation{Physics Division, Lawrence Livermore National Laboratory, Livermore, CA, 94550, USA}
\author{Phanish Suryanarayana}
\email[Email: ]{phanish.suryanarayana@ce.gatech.edu}
\affiliation{College of Engineering, Georgia Institute of Technology, Atlanta, GA 30332, USA}
\affiliation{College of Computing, Georgia Institute of Technology, Atlanta, GA 30332, USA}

\begin{abstract}
We present a GPU-accelerated version of the real-space SPARC electronic structure code for performing hybrid functional calculations in generalized Kohn-Sham density functional theory. In particular, we develop a batch variant of the recently formulated Kronecker product-based  linear solver for the simultaneous solution of multiple linear systems. We then develop a modular, math kernel based implementation for hybrid functionals on \texttt{NVIDIA} architectures, where computationally intensive operations are offloaded to the GPUs while the remaining workload is handled by the CPUs. Considering bulk and slab examples, we demonstrate that GPUs enable up to 8x speedup in node-hours and 80x in core-hours compared to CPU-only execution, reducing the time to solution on \texttt{V100} GPUs to around 300 seconds for a metallic system with over 6,000 electrons, and significantly reducing the computational resources required for a given wall time.
\end{abstract}

\maketitle

\section{Introduction}

Electronic structure calculations based on Kohn-Sham density functional theory (DFT) \cite{kohn1965self, hohenberg1964inhomogeneous} have become essential in materials and chemical sciences research, providing valuable insights and robust  predictive capabilities. The widespread adoption of DFT can be attributed to its balance of simplicity, generality, and high accuracy-to-cost ratio compared to other such ab initio methods \cite{becke2014dft, burke2012dft}. However, the cost of Kohn-Sham calculations increases rapidly with system size, limiting the range of systems that can be studied. These costs are further increased with the use of  advanced exchange-correlation functionals, particularly in ab initio molecular dynamics (AIMD) simulations, where hundreds of thousands  of Kohn-Sham solutions may be required to investigate certain properties or phenomena \cite{burke2012dft, kumar2024fly}.

The planewave pseudopotential method \cite{martin2020electronic} has been one of the most widely used solution approaches in Kohn-Sham DFT \cite{VASP, CASTEP, ABINIT, Espresso, CPMD, DFT++, gygi2008architecture, valiev2010nwchem}. Its accuracy stems from the Fourier basis, while its efficiency is a consequence of highly optimized Fast Fourier Transforms (FFTs). However, the  periodic nature of the basis restricts the planewave method to periodic boundary conditions, and its global nature hinders scalability on parallel computing platforms. These drawbacks have led to the development of alternative methods that employ systematically improvable, localized representations \cite{becke1989basis, chelikowsky1994finite, genovese2008daubechies, seitsonen1995real, white1989finite, iwata2010massively, tsuchida1995electronic, xu2018discrete, Phanish2011, Phanish2010, ONETEP, CONQUEST, MOTAMARRI2020106853, castro2006octopus, briggs1996real, fattebert1999finite, shimojo2001linear, ghosh2017sparc, arias1999wav, pask2005femeth, lin2012adaptive, xu2021sparc}. Among these, real-space finite-difference methods, which maximize computational locality and naturally accommodate Dirichlet as well as Bloch-periodic boundary conditions, are perhaps the most mature and widely used to date.  In particular, these methods have been successfully scaled to handle large systems containing up to a million atoms \cite{gavini2022roadmap, fattebert2016modeling}.

Hybrid density functionals, which are positioned on the fourth rung of Jacob's ladder, are orbital-dependent exchange-correlation functionals formulated within the framework of generalized Kohn-Sham DFT \cite{martin2020electronic, seidl1996generalized} that combine a portion of the nonlocal Hartree-Fock exact exchange energy with contributions from local/semilocal exchange-correlations. They can be broadly classified into the unscreened and screened/range-separated variants, the former more commonly used for isolated systems like  clusters and molecules whereas the latter are preferred for condensed matter systems, such as 3D bulk materials and surfaces.  Hybrid functionals provide superior predictive accuracy compared to local/semilocal approximations for a wide range of materials and properties, including lattice constant, bulk modulus, spin magnetic moment, ionization potential, atomization energy, proton affinity, bandgap, and heat of formation \cite{gill1992investigation, becke1993new, Becke1993DensityfunctionalTI, becke1996density, adamo1999toward, adamo1999accurate, garza2016predicting}. 

Hybrid functional calculations are significantly more computationally demanding than local/semilocal functionals, by up to two orders of magnitude. This has led to the development of various methods aimed at reducing the prefactor and/or scaling associated with these computations, in the context of planwewave methods \cite{lin2016adaptively, gygi2009compact, damle2015compressed, damle2017scdm, damle2017computing, hu2017projected, hu2017interpolative, mountjoy2017exact, ko2020enabling, ko2021enabling, ko2023high} as well as real-space finite-element \cite{subramanian2024tucker} and finite-difference \cite{jing2024efficient} methods more recently.  Given the processing power provided by Graphics Processing Units (GPUs) and their widespread availability on modern computers, performing computationally intensive operations on GPUs represents an attractive option to reduce  time to solution in electronic structure calculations \cite{walker2016electronic, gonze2016recent, genovese2009density, genovese2016wavelet, manninen2013applied, maintz2011speeding, hacene2012accelerating, jia2017gpu, andrade2012time, wilkinson2013porting, jia2013fast,romero2018performance, huhn2020gpu, das2022dft, sharma2023gpu}, and in hybrid functional calculations \cite{Ratcliff_2018, doi:10.1021/acs.jctc.9b01167, 10.1063/5.0208103} in particular.

SPARC \cite{xu2021sparc, zhang2024sparc} is a real-space finite-difference electronic structure code that can naturally accommodate Dirichlet, periodic, and Bloch-periodic boundary conditions, and their combinations, which allows for the accurate and efficient treatment of finite and semi-infinite as well as bulk 3D systems.  SPARC efficiently scales to large computational resources, leveraging thousands of processors in regular operation, which results in significant speedups --- often by an order of magnitude and more --- compared to planewave methods in the case of local, semilocal, and hybrid functionals, with increasing gains as the number of processors increases. A GPU-accelerated version of the SPARC real-space electronic structure code was recently developed \cite{sharma2023gpu}, achieving significant speedups over CPU-only execution and enabling substantial reductions in computational resource requirements for a given wall time. However, it was restricted to local/semilocal exchange-correlation functionals, which provides the motivation for the present work. 

In this work, we develop a GPU-accelerated version of SPARC for hybrid functional calculations. In particular, we introduce a batch variant of the Kronecker product-based linear solver \cite{jing2024efficient} for solving multiple systems simultaneously and implement a modular, math kernel based approach for hybrid functionals on \texttt{NVIDIA} GPUs. Benchmarking on bulk and slab systems shows up to 8x speedup in node-hours and 80x in core-hours compared to CPU-only execution, reducing solution time on \texttt{V100} GPUs to about 300 seconds for a metallic system with over 6,000 electrons, and significantly reducing computational resource requirements.

The remainder of this paper is organized as follows. In Sec.~\ref{Sec:Real-spaceKP}, we discuss the Kronecker product formalism for the solution of linear systems and its batch variant. In Sec.~\ref{Sec:GPU-hybridDFT}, we describe the GPU acceleration of hybrid functional calculations in the  SPARC electronic structure code. Next, we verify the performance of the GPU-accelerated implementation in Sec.~\ref{Sec:results}. Finally, we provide concluding remarks in Sec.~\ref{Sec:Conclusions}.


\section{Real space formulation \label{Sec:Real-spaceKP}}
Hybrid density functionals can be broadly classified as unscreened or screened/range-separated. The exact exchange operator and its screened variants take the form:
\begin{align}
V_{X}^{\sigma}\varphi_{n\bm k}^{\sigma}(\bm r) = -\sum_{m\bm q} w_{\bm q} g_{m\bm q}^{\sigma} \psi_{m\bm q}^{\sigma}(\bm r)  \phi_{m\bm q n\bm k}^{\sigma} (\bm{r}) \,, \label{Eq:Vx:Poisson} 
\end{align}
where $\psi$ are the orbitals, $g$ are the occupations, $w$ are the Brillouin zone weights, and $\varphi$ is any given function, with the quantities being indexed by the spin $\sigma \in \{\uparrow, \downarrow\}$, Brillouin zone wavevectors $\bm k$ and $\bm q$, and the band numbers $m$ and $n$. For unscreened calculations, $\phi$ can be written as the solution to the linear system \cite{jing2024efficient}:
\begin{equation}\label{Eq:Poisson}
- \frac{1}{4 \pi} \nabla^2 \phi_{m\bm q n\bm k}^{\sigma} (\mathbf{r}) =  \psi_{m\bm q}^{\sigma\ast}(\bm{r}) \varphi_{n\bm k}^{\sigma}(\bm{r}) \,,
\end{equation}
while for the screened counterparts \cite{jing2024efficient}:
\begin{align} \label{Eq:Poisson:SR}
-\frac{1}{4 \pi} \left( I - e^{ -\frac{\nabla^2}{16 \pi \omega^2} } \right)^{-1}  \nabla^2 \phi_{m\bm q n\bm k}^{\sigma}(\bm{r}) = \psi_{m\bm q}^{\sigma\ast}(\bm{r}) \varphi_{n\bm k}^{\sigma}(\bm{r}) \,,
\end{align}
both subject to Bloch boundary conditions at the  wavevector $\bm k - \bm q$ in the directions that the system is extended, and  Dirichlet boundary conditions in the directions of vacuum.  Above, $\omega$ is the screening parameter that determines the range separation for screened hybrid functionals. 

\subsection{Kronecker product formalism \label{Subsec:Kron:Original}}
Consider a real-space discretization on a uniform 3D grid containing $N = n_1n_2n_3$ points, with $n_1$, $n_2$, and $n_3$ grid points along the $x_1$, $x_2$, and $x_3$ directions, respectively. The solution to the linear systems in Eqs.~\ref{Eq:Poisson} and \ref{Eq:Poisson:SR}  can be written as \cite{jing2024efficient}:
\begin{align} \label{Eq:LS:Discrete:Solution}
\bX = f(\bL) \bB \,,
\end{align}
where $\bL$ is the discrete Laplacian matrix, $\bB$ is the right hand side vector, and the function $f$ is determined by the type of hybrid functional:
\begin{align} \label{Eq:fL}
 f(\bL) = \begin{cases} 
             -4 \pi \bL^{-1}  \quad  & \text{unscreened}\\
             -4 \pi \bL^{-1} (\mathbf{I} - e^{-\frac{\bL}{16 \pi \omega^2}}) \quad & \text{screened.}
             \end{cases} 
\end{align} 

On employing the Kronecker product decomposition of the Laplacian matrix, it follows that \cite{jing2024efficient}:
\begin{align} \label{Eq:FinalRepfL}
\bL  = (\bVx \otimes \bVy \otimes \bVz) \boldsymbol{\Lambda}  (\bVx^{* \rm T} \otimes \bVy^{* \rm T} \otimes \bVz^{* \rm T}) \,, 
\end{align}
where $\bVx$, $\bVy$, and $\bVz$ are the eigenvectors of the discrete second derivative operators along the $x_1$, $x_2$, and $x_3$ directions, respectively, and $\boldsymbol{\Lambda}$ is a diagonal matrix of the eigenvalues of $\bL$. Therefore, 
\begin{align} \label{Eq:FinalRepfL}
f(\bL)  = (\bVx \otimes \bVy \otimes \bVz)  f(\boldsymbol{\Lambda})  (\bVx^{* \rm T} \otimes \bVy^{* \rm T} \otimes \bVz^{* \rm T}) \,, 
\end{align}
using which the the solution to the linear systems can be written as \cite{jing2024efficient}:
\begin{subequations}
 \begin{align}
\bX  & = {\rm vec}_{n_3} \left[ \left( \bigwedge_{1 \le k\le n_3}  {\rm vec}_{n_2} (\bVx \widetilde{\bX}_k \bVy^{\rm T})  \right) \bVz^{\rm T} \right] \,,  \label{Eq:FinalSolution} \\
\widetilde{\bX} &  = f(\widetilde{\boldsymbol{\Lambda}}) \odot  {\rm vec}_{n_3} \left[ \left( \bigwedge_{1 \le k\le n_3}  {\rm vec}_{n_2} (\bVx^{* \rm T} \bB_k \bVy^{*}) \right) \bVz^{*} \right] \,. \label{Eq:LS:SolutionPart}
\end{align} 
\end{subequations}
where ${\rm vec}_{(.)}$ denotes the vectorization operator  along the subscripted  dimension, which converts the matrix to a column vector \cite{van2000ubiquitous, sharma2018real}; $\bigwedge$ represents the loop operator, which accumulates the different column vectors into a matrix \cite{sharma2018real}; $\widetilde{\bX}_k = \widetilde{\bX}(:,:,k)$ and $\bB_k = \bB(:,:,k)$ denote the frontal slices of the $\widetilde{\bX}$ and $\bB$ vectors, respectively, while within a multidimensional representation; $\widetilde{\boldsymbol{\Lambda}} = \text{diag} (\boldsymbol{\Lambda})$, where $\text{diag}(.)$ denotes the diagonal; and $\odot$ represents the Hadamard (element wise) product. In so doing, the solution of each linear system requires $4 n_3 + 2$ dense matrix-matrix multiplications.

\subsection{Batch Kronecker product formalism}
In the Kronecker product based formalism described above, each linear system is solved sequentially. In particular, the matrices involved in the matrix-matrix multiplications are small in size, whereby the performance gains (if any) from GPU acceleration are expected to be minimal. To overcome this,  we now exploit the fact that $f(\bL)$ is the same for each linear system at given Brillouin zone wavevectors $\mathbf{k}$ and $\mathbf{q}$ to develop a batch variant, i.e.,  multiple linear systems are solved simultaneously, wherein the matrices involved in the matrix-matrix multiplications are significantly larger, even in the limiting case  of a single linear system in each batch. 

Consider the solution to the linear systems written in block form as:
\begin{align} \label{Eq:LS:Discrete:Solution:Block}
\bY = f(\bL) \bC \,,
\end{align}
where 
\begin{subequations}
\begin{align}
\bY & = [\bX^{(1)}, \ldots, \bX^{(n_c)} ]  \,, \\
\bC & = [\bB^{(1)}, \ldots, \bB^{(n_c)} ]  \,,
\end{align}
\end{subequations}
the superscript used an index for the different linear systems, with $n_c$ representing the number of linear systems being solved simultaneously. In this case, the solution to the multiple linear systems can be written as:
\begin{subequations}
\begin{align}
\bY &=   \left[ \left( (\bVx \widetilde{\bY}_{[1]})_{[2]} \bVy^{\rm T})  \right)_{[3]} \bVz^{\rm T} \right]_{[4]}\,, \label{Eq:FinalSolution} \\
\widetilde{\bY}  & = f(\widetilde{\boldsymbol{\Lambda}})_{n_{\rm c}} \odot  \left[   \left( (\bVx^{* \rm T} \bC_{[1]})_{[2]} \bVy^{*}\right)   _{[3]}\bVz^{*} \right]_{[4]} \,, \label{Eq:LS:SolutionPart}
\end{align} 
\end{subequations}
where we employ four reorganizations of the data in evaluating each of the above expressions. First, 
\begin{align}
\bZ &\rightarrow \bZ_{[1]} \,, \nonumber \\
\bZ & = [\bZ^{(1)}, \ldots, \bZ^{(n_c)} ] \,,  \\ 
\bZ_{[1]} & = [\tilde{\bZ}^{(1)}, \ldots, \tilde{\bZ}^{(n_c)} ]  \,,  \,\,
\tilde{\bZ}^{(j)} & = [\tilde{\bZ}^{(j)}_1,  \ldots, \tilde{\bZ}^{(j)}_{n_3} ]  \,, \nonumber 
\end{align}
where $\tilde{\bZ}^{(j)}_k = \bZ^{(j)}(:,:,k)$. Second, 
\begin{align}
\bZ_{[1]} &\rightarrow \bZ_{[2]} \,, \nonumber \\
\bZ_{[1]} & = [\tilde{\bZ}^{(1)},\ldots, \tilde{\bZ}^{(n_c)} ]  \,, \,\,
\tilde{\bZ}^{(j)}  = 
\begin{bmatrix}
\tilde{\bZ}^{(j)}_1 \\
\vdots \\
\tilde{\bZ}^{(j)}_{n_3}
\end{bmatrix}  \,, \nonumber \\
\bZ_{[2]} & = 
\begin{bmatrix}
\tilde{\bZ}^{(1)}\\ 
\vdots \\
\tilde{\bZ}^{(n_c)} 
\end{bmatrix} \,.
\end{align}
Third,
\begin{align}
\bZ_{[2]} &\rightarrow \bZ_{[3]} \,, \nonumber \\ 
\bZ_{[2]} & = 
\begin{bmatrix}
\tilde{\bZ}^{(1)}\\ 
\vdots \\
\tilde{\bZ}^{(n_c)} 
\end{bmatrix} \,, \,\,
\tilde{\bZ}^{(j)}  = 
\begin{bmatrix}
\tilde{\bZ}^{(j)}_1 \\
\vdots \\
\tilde{\bZ}^{(j)}_{n_3}
\end{bmatrix}  \,, \\  
\bZ_{[3]} & = 
\begin{bmatrix}
\tilde{\tilde{\bZ}}^{(1)}\\ 
\vdots \\
\tilde{\tilde{\bZ}}^{(n_c)} 
\end{bmatrix} \,,  \,\,
\tilde{\tilde{\bZ}}^{(j)} = [{\rm vec}_{n_2} (\tilde{\bZ}_1^{(j)}) ,  \ldots, {\rm vec}_{n_2} (\tilde{\bZ}_{n_3}^{(j)})  ] \,. \nonumber 
\end{align}
Fourth,
\begin{align}
\bZ_{[3]} & = 
\begin{bmatrix}
\tilde{\tilde{\bZ}}^{(1)}\\ 
\vdots \\
\tilde{\tilde{\bZ}}^{(n_c)} 
\end{bmatrix} \,, \,\, \tilde{\tilde{\bZ}}^{(j)} = [\tilde{\tilde{\bZ}}^{(j)}_1,  \ldots, \tilde{\tilde{\bZ}}^{(j)}_{n_3} ] \,, \nonumber \\
\bZ_{[4]} & = [{\rm vec}_{n_3} (\tilde{\tilde{\bZ}}^{(j)}_1),  \ldots, {\rm vec}_{n_3}(\tilde{\tilde{\bZ}}^{(j)}_{n_3}) ] \,.
\end{align}
In addition, 
\begin{align}
f(\widetilde{\boldsymbol{\Lambda}})_{n_{\rm c}} = [f(\widetilde{\boldsymbol{\Lambda}}), \ldots, f(\widetilde{\boldsymbol{\Lambda}})] \,,
\end{align}
a matrix with $n_c$ columns. In so doing, the solution of each linear system requires $6/n_c$ dense matrix-matrix multiplications. Notably, even in the case of $n_c=1$, the batch formalism requires only 6 matrix-matrix multiplications, relative to the $4 n_3 + 2$ multiplications in the original formalism (Section~\ref{Subsec:Kron:Original}). 

\begin{figure}[htbp!]
\centering
  \includegraphics[width=0.45\textwidth]{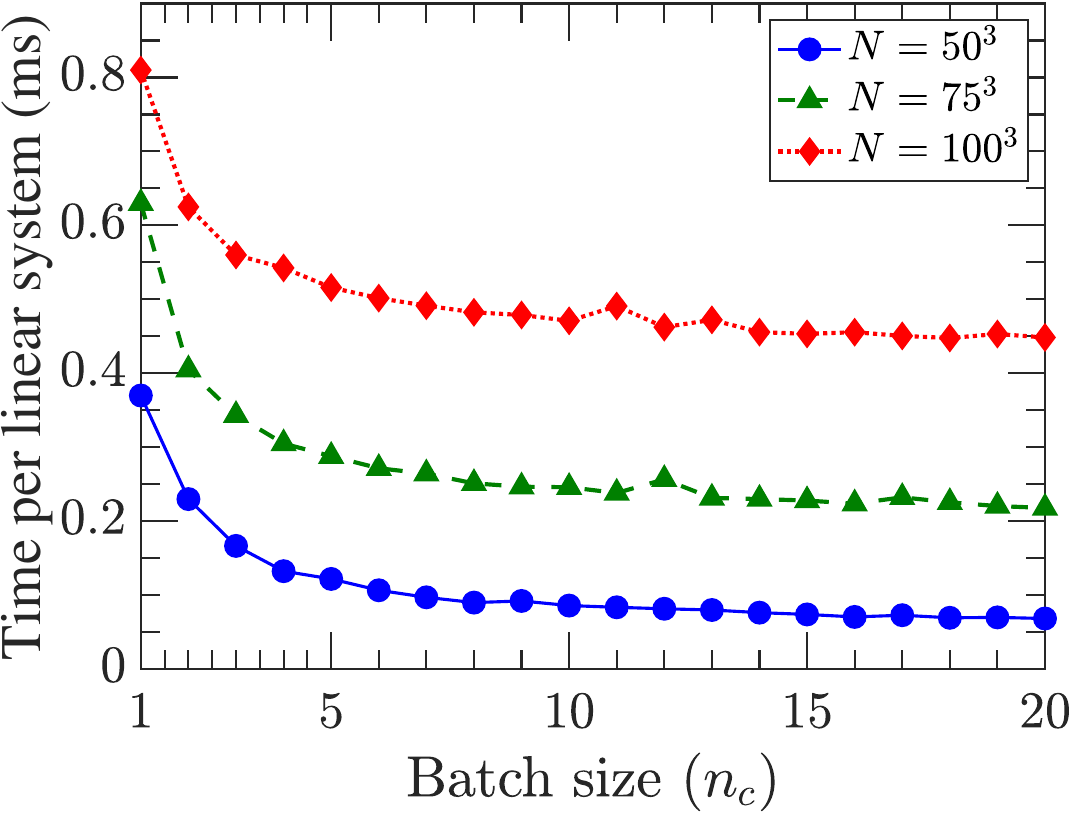}
  \caption{Time to solution per linear system (Poisson equation) as a function of the batch size for the batch Kronecker product formalism on the \texttt{NVIDIA V100} GPU. In all instances, the timings are averaged over 50 runs. \label{Fig:mc}}
\end{figure}

In Fig.~\ref{Fig:mc}, we present the time to solution per linear system as a function of the batch size for the batch Kronecker product formalism on the \texttt{NVIDIA} V100 GPU. In particular, we consider the Poisson equation, which is solved on cubical domains with 50, 75, 100 grid points in each direction, while holding the grid spacing constant.. We observe an increase in the speed with batch size $n_c$, stagnating at batch sizes of $n_c = \mathcal{O}(20)$, achieving speedups of 2 to 4x relative to $n_c=1$, with larger speedups for smaller number of grid points. In view of this, we will consider a batch size of $n_c=20$ for the simulations in this work.

\begin{figure}[htbp!]
\centering
  \includegraphics[width=0.45\textwidth]{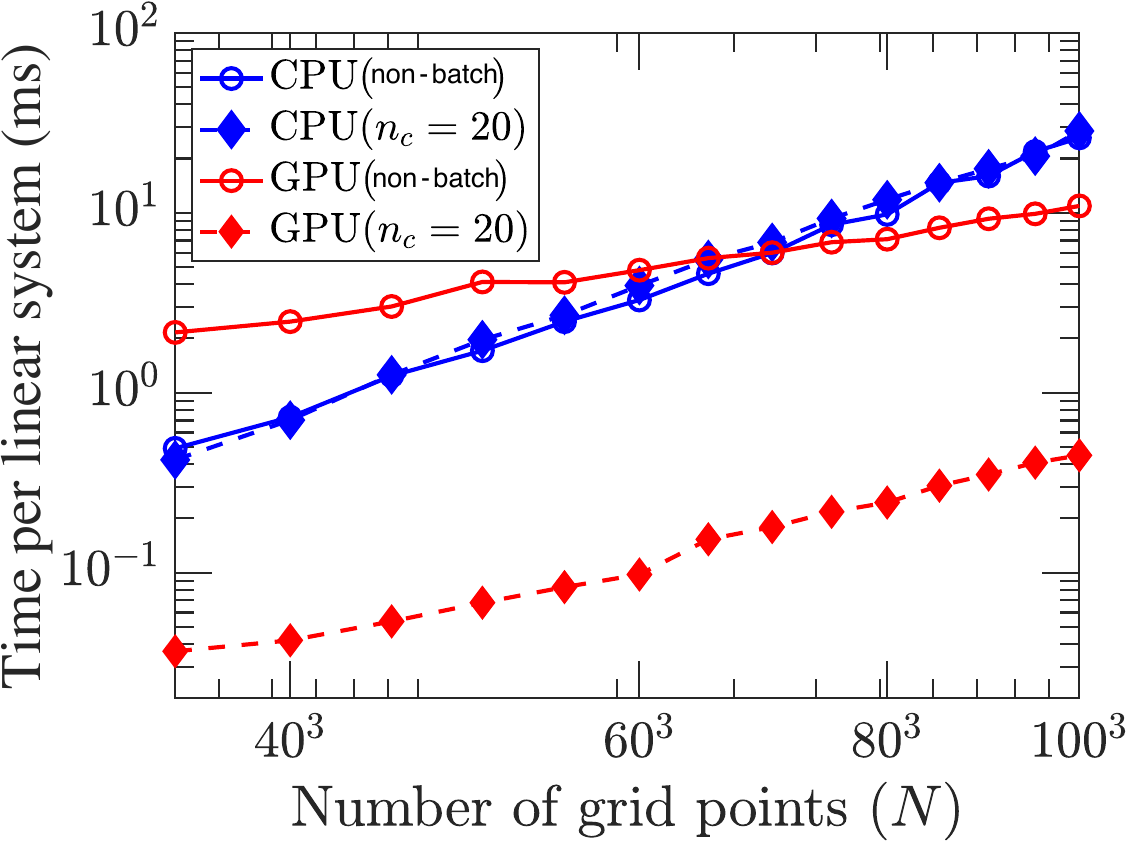}
  \caption{Time to solution per linear system (Poisson equation) as a function of the number of grid points for the batch Kronecker product formalism on both CPU and \texttt{NVIDIA V100} GPU, compared to the original (non-batch) Kronecker product formalism on both CPU and GPU. In all instances, the timings are averaged over 50 runs. \label{Fig:compare}}
\end{figure}

In Fig.~\ref{Fig:compare}, we present the time to solution per linear system (Poisson equation) as a function of the number of grid points for the batch Kronecker product formalism on both CPU and  \texttt{NVIDIA V100} GPU, compared to the original Kronecker product formalism on both CPU and GPU. We observe that the original Kronecker product formalism has comparable performance on CPU and GPU, which can be attributed to the relatively small size of the matrices involved in the dense matrix-matrix multiplications. We also observe that the batch Kronecker product formalism on GPU is significantly faster than CPU,  with speedups of 24, 33, and 51x for the systems with $50^3$, $75^3$, $100^3$ grid points, respectively. Note that the speedup of the batch version relative to the original formalism on GPU is a consequence of solving multiple linear systems simultaneously, i.e., $n_c=20$ (Fig.~\ref{Fig:mc}), as well as the batch formalism for $n_c=1$ itself involving matrix-matrix multiplications that are significantly fewer in number and therefore of significantly larger size relative to the original formalism (Section~\ref{Subsec:Kron:Original}). Note that though it is possible to achieve the same by using libraries that provide batch functions, e.g., \texttt{cblas\_dgemm\_batch} and \texttt{cblas\_zgemm\_batch} in intel MKL, we still develop and implement the above batch Kronecker product based formalism to maximize portability of the code. Indeed, we have verified that the performance of the developed code is competitive with such libraries.

\section{GPU acceleration \label{Sec:GPU-hybridDFT}} 

Hybrid functional simulations in the SPARC electronic structure code \cite{xu2021sparc, zhang2024sparc} proceed as follows. In each calculation, the electronic ground state corresponding to the PBE \cite{perdew1996generalized} exchange-correlation functional is first determined, with the resulting orbitals and density serving as an initial guess for the hybrid functional calculation. An outer fixed-point iteration is employed with respect to the exact exchange operator \cite{lin2016adaptively}, in addition to the standard inner fixed-point iteration with respect to the density/potential, commonly referred to as the self-consistent field (SCF) method \cite{martin2020electronic}. In particular, the adaptively compressed exchange (ACE) operator method \cite{lin2016adaptively} is used in conjunction with the Chebyshev filtered subspace iteration (CheFSI) \cite{zhou2006parallel, zhou2006self}, which is accelerated via the restarted variant of the preconditioned Periodic Pulay mixing scheme \cite{banerjee2016periodic}. The Poisson problem for the electrostatics \cite{ghosh2017sparc, Phanish2010} is solved using the alternating Anderson-Richardson (AAR) method \cite{suryanarayana2019alternating}.

The overall parallelization scheme adopted in SPARC for electronic structure calculations is as follows. In each inner SCF iteration, the  electrostatic Poisson equation is solved on a Cartesian topology formed by embedding a three-dimensional processor grid into the \texttt{MPI\_COMM\_WORLD} communicator. The mixing of the density/potential is also performed on this topology. The linear eigenproblem is solved in the eigensolver topology, i.e., the Kohn-Sham orbitals are partitioned using this topology. The eigensolver topology  consists of smaller Cartesian topologies, created by partitioning the \texttt{MPI\_COMM\_WORLD} communicator into two spin groups, which are then subdivided into $p_1$ Brillouin zone integration (k-point) groups, further divided into $p_2$ band groups, and finally, each band group is embedded with a Cartesian topology mapped to  $p_3$ processors.  

In this work, building on recent efforts to accelerate local/semilocal DFT calculations in SPARC using GPUs \cite{sharma2023gpu}, we develop a GPU-accelerated version for hybrid functional calculations. In so doing, the GPU-accelerated version for local/semilocal functionals has been extended to include domain decomposition. In what follows, we describe how the key computational kernels in hybrid calculations --- specifically, the construction and application of the ACE operator --- are accelerated on \texttt{NVIDIA} GPUs using the \texttt{cuBLAS} and \texttt{cuSOLVER} libraries within the CUDA parallel programming platform. We set the CPU-thread-to-GPU ratio to 1 --- each CPU rank to be directly assigned to the corresponding GPU rank --- which ensures optimal load balancing, minimizes PCI bus transactions, and provides the most efficient performance. The data transfers between GPUs are handled using the \texttt{NVIDIA} Collective Communications Library (\texttt{NCCL}), while transfers between CPUs are managed using the Message Passing Interface (\texttt{MPI}). The data transfers from CPU to GPU and GPU to CPU are performed using the \texttt{cublasSetVector}/\texttt{cublasSetMatrix} and \texttt{cublasGetVector}/\texttt{cublasGetMatrix} routines, respectively. Note that for isolated systems or $\Gamma$-point calculations, real-valued computations are performed, while for systems with Brillouin zone integration, complex-valued computations are performed, with all operations carried out in double-precision arithmetic.

The implementation developed in this work supports both spin-unpolarized and spin-polarized calculations. For simplicity, we omit spin considerations in the following discussion. Specifically, we focus on one spin group, with the implementation for the other spin group following the same approach, as the eigenproblems for different spins are essentially identical and independent.
\subsection{ACE operator construction}

 The ACE operator in discrete form can be written as:
\begin{align}
\widetilde{V}_{\bm k} = \bm \xi_{\bm k} \bm \xi_{\bm k}^{\rm T} \,,
\end{align}
where 
\begin{align}
{\bm \xi}_{\bm k} = \mathbf{W}_{\bm k} \mathbf{R}_{\bm k} ^{-1}  \,.
    \label{Eq:xi}
\end{align}
Above,
\begin{subequations}
\begin{align}
\mathbf{W}_{\bm k} & = [\mathbf{W}_{\bm k}^{(1)}, \mathbf{W}_{\bm k}^{(2)}, \ldots, \mathbf{W}_{\bm k}^{(N_s)}] \,, \\
\mathbf{W}_{\bm k}^{(n)}(\bm r) & = -\sum_{m\bm q} w_{\bm q} g_{m\bm q} \psi_{m\bm q}(\bm r) \phi_{m\bm q n\bm k} (\bm{r}) \,, \label{Eq:W}
\end{align}
\end{subequations}
where $N_s$ is the number of occupied orbitals. In addition, $\mathbf{R}_{\bm k}$ is a lower triangular matrix that represents the Cholesky factor of the matrix
\begin{align}
\mathbf{M}_{\bm k} = {\bm \Psi}_{\bm k}^{\rm T} \mathbf{W}_{\bm k} \,,
\end{align}
where ${\bm \Psi}_{\bm k}$ represents the collection of orbitals.  

\begin{figure}[htbp!]
\centering
  \includegraphics[width=0.4\textwidth]{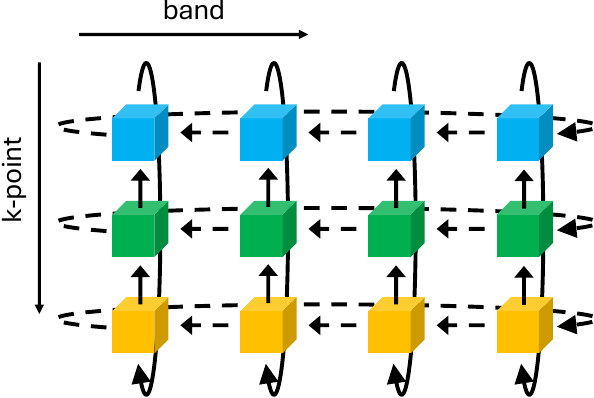}
  \caption{Illustration of the two-level ring communication pattern used to transfer orbitals between CPUs, specifically for the case of 4 band groups and 3 k-point groups. \label{Fig:RingComm}}
\end{figure}

The construction of the ACE operator thus requires the solution of the linear systems described by Eqs.~\ref{Eq:Poisson} and \ref{Eq:Poisson:SR} in the case of  unscreened and screened hybrid functional calculations, respectively.  Due to the large number of linear systems to be solved, they are solved sequentially using the batch Kronecker product-based formalism.  This requires forming the product of each orbital with itself and every other orbital in the same spin group, including those corresponding to different k-point groups. To do so, the orbitals are transferred between CPUs using a two-level ring communication pattern, as illustrated in Fig.~\ref{Fig:RingComm}. In particular, a staggered communication pattern is adopted, with ring communication occurring between band groups at the inner level and between k-point groups at the outer level. 

The local part of the orbitals initially assigned to each processor, along with those made available during each cycle of the ring communication, are transferred from the CPU to the corresponding GPU. The local part of the right-hand side vectors, corresponding to the orbitals on each GPU, are then computed. After splitting all the right-hand sides available in each band group into groups of $n_c p_3$, they are redistributed so that $n_c$ right-hand sides are assigned to each of the $p_3$ GPUs within each band group:
\begin{align}
\begin{array}{r}
 \text{\scriptsize GPU}_{1} \{ \\
\vspace{3mm}  \\
\text{\scriptsize GPU}_{\rm p_3} \{ \\
\end{array} 
\hspace{-2.5mm}
\begin{bmatrix}
\bC_1^{(1)} & \ldots & \bC_{1}^{(p_3)}\\
\vdots &\ddots & \vdots \\
\bC_{p_3}^{(1)} & \ldots & \bC_{p_3}^{(p_3)} \\
\end{bmatrix} 
\rightarrow
\begin{bmatrix}
\bC_1^{(1)} & \ldots &  \bC_{p_3}^{(1)} \\
\vdots &\ddots & \vdots \\
\bC_{1}^{(p_3)} & \ldots & \bC_{p_3}^{(p_3)} \\
\end{bmatrix} \hspace{-2.5mm}
\begin{array}{l}
\} \text{\scriptsize GPU}_{1} \\
\vspace{3mm}  \\
\} \text{\scriptsize GPU}_{\rm p_3} \\
\end{array}  \,.
\end{align}
Once these linear systems have been solved using the batch Kronecker product solver, the solutions are transferred back to the original layout:
\begin{align}
\begin{array}{r}
 \text{\scriptsize GPU}_{1} \{ \\
\vspace{3mm}  \\
\text{\scriptsize GPU}_{\rm p_3} \{ \\
\end{array} 
\hspace{-2.5mm}
\begin{bmatrix}
\bY_1^{(1)} & \ldots &  \bY_{p_3}^{(1)} \\
\vdots &\ddots & \vdots \\
\bY_{1}^{(p_3)} & \ldots & \bY_{p_3}^{(p_3)} \\
\end{bmatrix} 
\rightarrow
\begin{bmatrix}
\bY_1^{(1)} & \ldots & \bY_{1}^{(p_3)}\\
\vdots &\ddots & \vdots \\
\bY_{p_3}^{(1)} & \ldots & \bY_{p_3}^{(p_3)} \\
\end{bmatrix} 
\hspace{-2.5mm}
\begin{array}{l}
\} \text{\scriptsize GPU}_{1} \\
\vspace{3mm}  \\
\} \text{\scriptsize GPU}_{\rm p_3} \\
\end{array}  \,.
\end{align}
To enable the above communication between GPUs, customized \texttt{NCCL\_Neighbor\_alltoallv} and \texttt{NCCL\_Neighbor\_alltoallv\_dist\_graph} routines are created by combining point-to-point \texttt{ncclSend} and \texttt{ncclRecv} targeting all neighbors, along with \texttt{ncclGroupStart} and \texttt{ncclGroupEnd} to improve efficiency and performance. The procedure is repeated until all linear systems that can be formed during each cycle of the ring communication have been solved, with the solutions used to compute the corresponding components of the matrix entries of $\mathbf{W}_{\bm k}$. Note that we will henceforth drop the index $\bk$, as there is no further communication or operations between the different k-point groups. 

Once the two-level ring communication has been completed,  $\mathbf{W}$ is available with the following partitioning:
\begin{align}
\mathbf{W} := 
\begin{bmatrix}
\bW_1^{(1)} & \ldots &  \bW_{1}^{(p_2)} \\ 
\vdots &\ddots & \vdots \\
\bW^{(1)}_{p_3} & \ldots & \bW^{(p_2)}_{p_3}\\
\end{bmatrix} \,.
\end{align}
Then, the \texttt{ncclAllReduce} routine is used such that the corresponding GPUs in each  band group get access to all the columns of $\mathbf{W}$, i.e., still with domain decomposition:
\begin{align}
\begin{bmatrix}
\bW_1^{(1)} & \ldots &  \bW_{1}^{(p_2)} \\ 
\vdots &\ddots & \vdots \\
\bW^{(1)}_{p_3} & \ldots & \bW^{(p_2)}_{p_3}\\
\end{bmatrix} 
\xrightarrow{\texttt{ncclAllReduce}}
\begin{bmatrix}
\bW_1 \\ 
\vdots \\
\bW_{p_3}
\end{bmatrix} \,, \\
\bW_i = \underbrace{
[
\underbrace{\bW_{i}^{(1)}}_{\text{GPU}_i^{(1)}} + 
\ldots +
\underbrace{\bW_{i}^{(p_2)}}_{\text{GPU}^{\rm (p_2)}_i}
] }_{\texttt{ncclAllReduce}} \,, \nonumber  \\
\bW_{i}^{(j)} \in {\rm GPU}_{i}^{(j)} \,, i \in \{1, \ldots, p_3\} \,, \, j \in \{1, \ldots, p_2\} \,, \nonumber \\
\bW_{i} \in {\rm GPU}_i^{(j)} \,, \, j \in \{1, \ldots, p_2\} \,. \nonumber 
\end{align}

The matrix $\bM$ is then calculated as:
\begin{align}
\bM := 
\begin{bmatrix}
\bM^{(1)} \\
\vdots \\
\bM^{(p_2)} 
\end{bmatrix} 
=
\begin{bmatrix}
\bpsi_{1}^{(1)} & \ldots &  \bpsi_{1}^{(p_2)} \\ 
\vdots &\ddots & \vdots \\
\bpsi^{(1)}_{(p_3)} & \ldots & \bpsi^{(p_2)}_{(p_3)}\\
\end{bmatrix}^{\rm T} 
\begin{bmatrix}
\bW_{1} \\
\vdots \\
\bW_{p_3} 
\end{bmatrix} \,,
 \\
\bM^{(j)} = \underbrace{
[
\underbrace{\bpsi_{1}^{(j)^{\rm T}} \bW_{1}}_{\text{GPU}_1^{(j)}} + 
\ldots +
\underbrace{\bpsi_{p_3}^{(j)^{\rm T}} \bW_{p_3}}_{\text{GPU}_{\rm p_3}^{(j)}}
] }_{\texttt{ncclAllReduce}} \,, \nonumber  \\
\bM^{(j)} \in {\rm GPU}_{i}^{(j)} \,, \, i \in \{1, \ldots, p_3\}  \,, \nonumber 
\end{align}
where the \texttt{cublasDgemm}/\texttt{cublasZgemm} routine is used for the matrix-matrix multiplications in the case of real/complex-valued computations. Once matrix $\bM$ has been computed, the \texttt{ncclAllReduce} routine is used over the band groups to ensure that the full matrix $\bM$ is available on each GPU. The matrix $\mathbf{R}$ is then calculated on each GPU by performing the Cholesky factorization of $\bM$  using the \texttt{cusolver\_dpotrf}/\texttt{cusolver\_zpotrf}  routine in the case of real/complex-valued computations. 

Finally, the matrix $\bm \xi$ is computed:
\begin{align}
\bxi 
:= 
\begin{bmatrix}
\bxi_1 \\
\vdots \\
\bxi_{p_3}
\end{bmatrix} 
=
\begin{bmatrix}
\bW_1 \\
\vdots \\
\bW_{p_3} 
\end{bmatrix}
\mathbf{R}^{-1} 
=
\begin{bmatrix}
\bW_1 \mathbf{R}^{-1}  \\
\vdots \\
\bW_{p_3}  \mathbf{R}^{-1} 
\end{bmatrix} \,, \\
\bxi_{i}, \bW_{i} \in  {\rm GPU}_i^{(j)} \,, \, j \in \{1, \ldots, p_2\}  \,, \nonumber 
\end{align}
where the \texttt{cublasDtrsm}/\texttt{cublasZtrsm} routine is used for the matrix-matrix multiplications in the case of real/complex-valued computations.

\subsection{ACE operator application}
The application of the ACE operator on any set of trial orbitals $\bPhi$:
\begin{align}
\mathbf{V} \bPhi = \bxi \bxi^{\rm T} \bPhi \,,
\end{align}
is computed in two steps. First,  we evaluate:
\begin{align}
\bchi & := 
\begin{bmatrix}
\bchi^{(1)} & \hdots & \bchi^{(p_2)}
\end{bmatrix} \nonumber \\
&= \bxi^{\rm T} \bPhi  =
\begin{bmatrix}
\bxi_1^{\rm T} & \hdots & \bxi_{p_3}^{\rm T} 
\end{bmatrix}
\begin{bmatrix}
\bpsi_{1}^{(1)} & \ldots &  \bpsi_{1}^{(p_2)} \\ 
\vdots &\ddots & \vdots \\
\bpsi^{(1)}_{(p_3)} & \ldots & \bpsi^{(p_2)}_{(p_3)}\\
\end{bmatrix} \,, \\
& \bchi^{(j)}  = \underbrace{
[
\underbrace{\bxi_{1}^{\rm T} \bpsi_{1}^{j}}_{\text{GPU}_1^j} + 
\ldots +
\underbrace{\bxi_{p_3}^{\rm T} \bpsi_{p_3}^{j}}_{\text{GPU}_{\rm p_3}^j}
] }_{\texttt{ncclAllReduce}} \,, \nonumber \\
& \bchi^{(j)} \in {\rm GPU}^{(j)}_{i} \,, \, i \in \{1, \ldots, p_3\} \,. \nonumber 
\end{align}
Next, we evaluate:
\begin{align}
 \bxi \bchi 
& =
\begin{bmatrix}
\bxi_1 \\
\vdots \\
\bxi_{p_3} 
\end{bmatrix} 
\begin{bmatrix}
\bchi^{(1)} \ldots \bchi^{(p_2)} 
\end{bmatrix} \nonumber \\
& = 
\begin{bmatrix}
\bxi_1 \chi^{(1)} & \ldots &  \bxi_1 \chi^{(p_2)} \\ 
\vdots &\ddots & \vdots \\
\bxi_{p_3} \bchi^{(1)} & \ldots & \bxi_{p_3} \bchi^{(p_2)} \\
\end{bmatrix}  \,, \\
& \bxi_i \chi^{(j)} \in { \rm GPU}_i^{(j)}\,, \,  i \in \{1, \ldots, p_3\},  j \in \{1, \ldots, p_2\} \,. \nonumber 
\end{align}

\section{Results and Discussion} \label{Sec:results}
We now assess the performance of the GPU-accelerated hybrid functional implementation in SPARC using representative examples: bulk molybdenum (Mo) and an 8-layer (100) slab of titanium dioxide (TiO$_2$) \cite{sahoo2022ab}. In particular, we perform isokinetic ensemble (NVK) ab initio molecular dynamics (AIMD) simulations with a Gaussian thermostat \cite{minary2003algorithms} at temperatures of 3000 K for Mo and 300 K for TiO$_2$, using time steps of 1 and 2 fs, respectively. We consider 128-, 250-, and 432-atom cells of Mo with the screened HSE exchange-correlation functional \cite{heyd2003hybrid, krukau2006influence} and $\Gamma$-point for  Brillouin zone integration; and 24-, 96-, and 384-atom cells of TiO$_2$  with the unscreened  PBE0 exchange-correlation functional \cite{Becke1993DensityfunctionalTI, perdew1996rationale} and Monkhorst-Pack \cite{monkhorst1976special} grids of $4 \times 4$, $2 \times 2$, and $1 \times 1$ for Brillouin zone integration, respectively. In all cases, we use ONCV pseudopotentials \cite{hamann2013optimized} with nonlinear core corrections from the SPMS table \cite{shojaei2023soft}, which includes 14, 12, and 6 electrons in valence for Mo, Ti, and O, respectively. 

The number of orbitals chosen for the Mo systems, Mo$_{128}$, Mo$_{250}$, and Mo$_{432}$, are $N_s = 1080$, $2105$, and $3633$, respectively; and for the TiO$_2$ systems (TiO$_2$)$_{8}$, (TiO$_2$)$_{32}$, and (TiO$_2$)$_{128}$, are $N_s = 120$, $465$, and $1848$, respectively, as automatically determined by SPARC. The grid spacings used for the Mo and TiO$_2$ systems are 0.358 and 0.3 bohr, respectively, corresponding to $N_d = 67\times 67 \times 67$, $83\times 83 \times 83$, and $100\times 100 \times 100$ finite-difference nodes for the Mo$_{128}$, Mo$_{250}$, and Mo$_{432}$, respectively, and $N_d = 30\times 30 \times 119$, $59\times 59 \times 119$, and $117\times 117 \times 119$ for the (TiO$_2$)$_{8}$, (TiO$_2$)$_{32}$, and (TiO$_2$)$_{128}$ systems, respectively. All numerical parameters in the DFT calculations, including grid spacing and SCF tolerances, are chosen to achieve a chemical accuracy of  $10^{-3}$ ha/atom in the energy.

We perform all simulations on the \texttt{Lassen} supercomputer at Lawrence Livermore National Laboratory \cite{LLNLwebMachines}. Each computational node is equipped with 4 \texttt{NVIDIA Volta V100} GPUs, each having 16 GB of memory, and 40 \texttt{IBM POWER9} CPU cores with a total of 256 GB of memory. In CPU-only runs, all 40 CPU cores per node are used, with one MPI thread per core. For GPU-accelerated runs, 4 CPU cores and 4 GPUs are allocated per node, with one MPI thread per GPU. The timing data is collected after around 10 AIMD steps, once the wall time per step has stabilized. In particular, the timings correspond to 2 PBE SCF iterations, 2 HSE outer loops, and 4 HSE inner iterations for the Mo systems; and 2 PBE SCF iterations, 2 PBE0 outer loops, and 5 PBE0 inner iterations for the TiO$_2$ systems, as well as the calculation of the Hellmann-Feynman atomic forces in all cases. 

\begin{figure*}[htbp!]
\centering
\subfloat[Bulk molybdenum]{\includegraphics[keepaspectratio=true,width=0.42\textwidth]{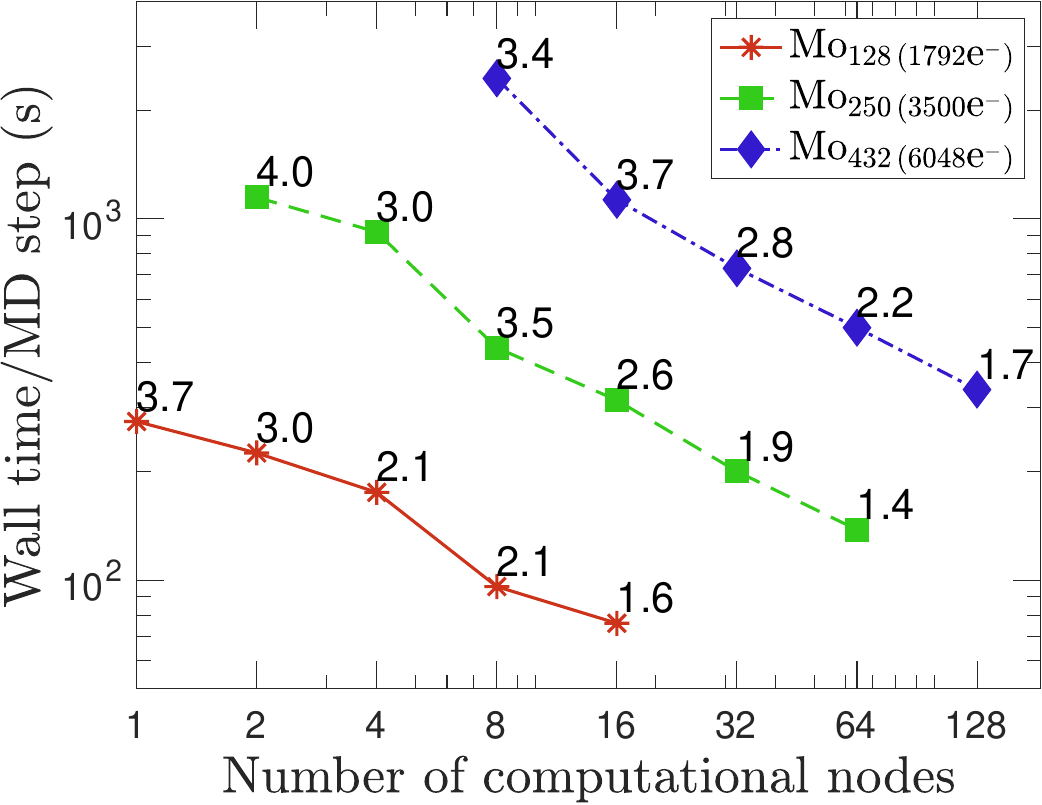} \label{Fig:Mo_scaling} } \hspace{8mm}
\subfloat[(001) Titanium dioxide slab]{\includegraphics[keepaspectratio=true,width=0.42\textwidth]{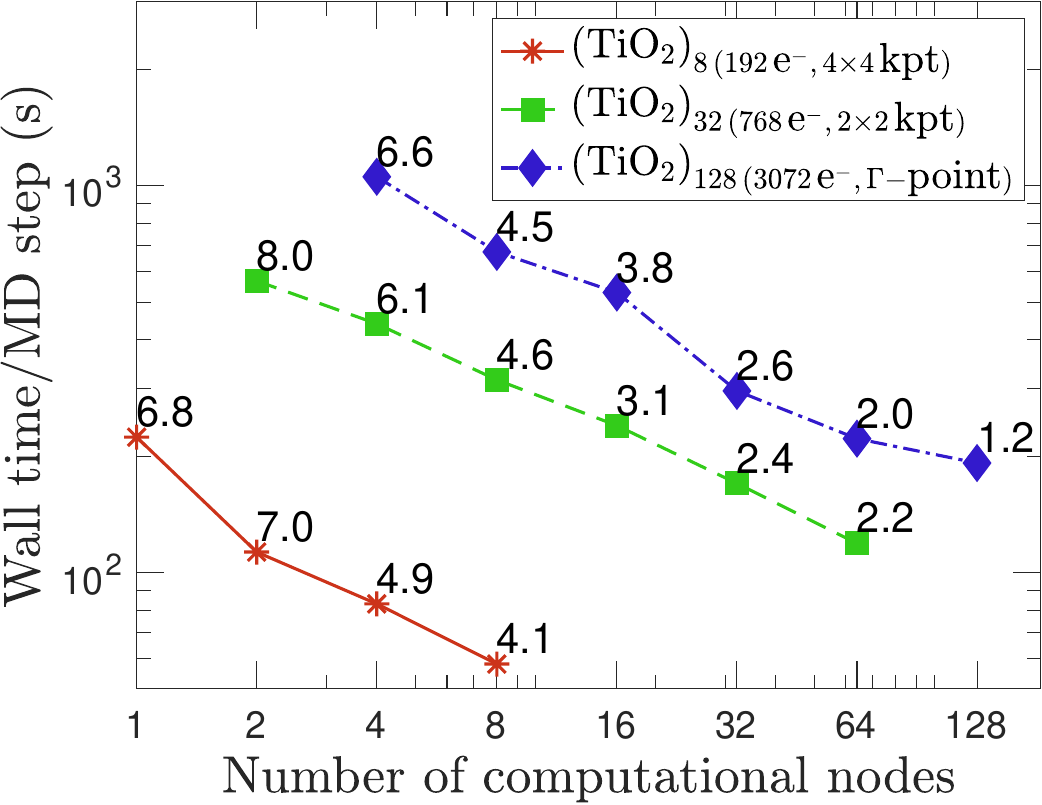} \label{Fig:TiO2_scaling} }
\caption{\label{Fig:Scaling} Strong scaling of the MD step time for GPU-accelerated hybrid functional calculations in SPARC on the \texttt{Lassen} supercomputer \cite{LLNLwebMachines}, where each computational node has 4 \texttt{NVIDIA V100} GPUs and 40 CPU cores. The timings correspond to using 4 GPUs and 4 CPU threads on each computational node. The number displayed next to each marker represents the speedup in time to solution relative to CPU-only execution, wherein  all 40  CPU cores on each computational node are utilized, i.e., the speedups in core hours are a factor of 10 larger.}
\end{figure*}

In Fig.~\ref{Fig:Scaling}, we present the strong scaling results so obtained for the selected Mo and TiO$_2$ systems.  In particular, the total wall time per MD step is plotted as a function of the number of computational nodes. We observe that the GPU implementation exhibits good parallel scaling, with a steady reduction in solution time as the number of nodes increases. The GPU-accelerated execution provides considerable speedup compared to CPU-only execution, achieving a maximum speedup of 3.7x, 4.0x, and 3.4x for Mo$_{128}$, Mo$_{250}$, and Mo$_{432}$, respectively; and 7.0x, 8.0x, and 6.6x for (TiO$_2$)$_{8}$, (TiO$_2$)$_{32}$, and (TiO$_2$)$_{128}$, respectively. Furthermore, the minimum MD step times are 76, 142, and 337 seconds for Mo$_{128}$, Mo$_{250}$, and Mo$_{432}$; and 58, 119, and 192 seconds for (TiO$_2$)$_{8}$, (TiO$_2$)$_{32}$, and (TiO$_2$)$_{128}$, respectively, demonstrating the  attractiveness of GPU-acceleration for AIMD with hybrid functionals.  The results clearly show that speedups are inversely correlated with the number of computational nodes and directly correlated with the problem size, similar to the observations for local/semilocal exchange-correlation functionals \cite{sharma2023gpu}. This can be attributed to two main factors. First, GPU-GPU neighbor communication via \texttt{NCCL} performs similarly to CPU-CPU communication using \texttt{MPI}, forming a significant part of the total wall time and becoming a bottleneck at larger number of computational nodes. Second, and perhaps more importantly, GPUs achieve greater acceleration when utilization is high, as they can process larger volumes of data and computational tasks. To verify this, we ran the Mo$_{432}$ simulation with a grid spacing of 0.22 bohr, which can be interpreted as either choosing a harder pseudopotential or targeting higher accuracy. In this case, we observed a speedup of 3.3x and a wall time of 1805 seconds on 64 computational nodes. In comparison, the corresponding numbers for 0.358 bohr grid spacing (Fig.~\ref{Fig:Scaling}) were 2.2x speedup and 488 seconds, respectively. Even though the number of finite-difference nodes increased by a factor of 4.3x, the wall time only increased by a factor of 3.4x, despite the Chebyshev polynomial degree rising from 22 to 33. Overall, the largest speedups are achieved with the smallest resources, making the reduction in wall time especially valuable for production runs, where resources are typically limited.

\begin{figure*}[htbp!]
\centering
\subfloat[Bulk molybdenum]{\includegraphics[keepaspectratio=true,width=0.42\textwidth]{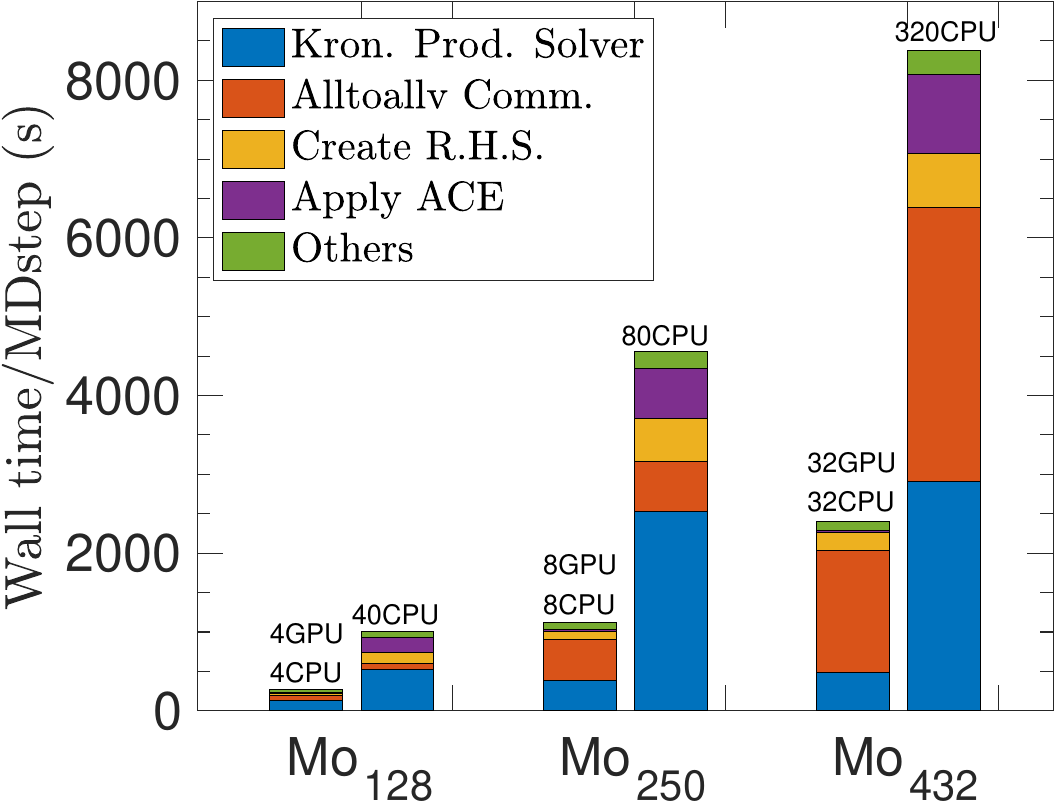} \label{Fig:Mo_timesplit_min} } \hspace{8mm}
\subfloat[(001) Titanium dioxide slab]{\includegraphics[keepaspectratio=true,width=0.42\textwidth]{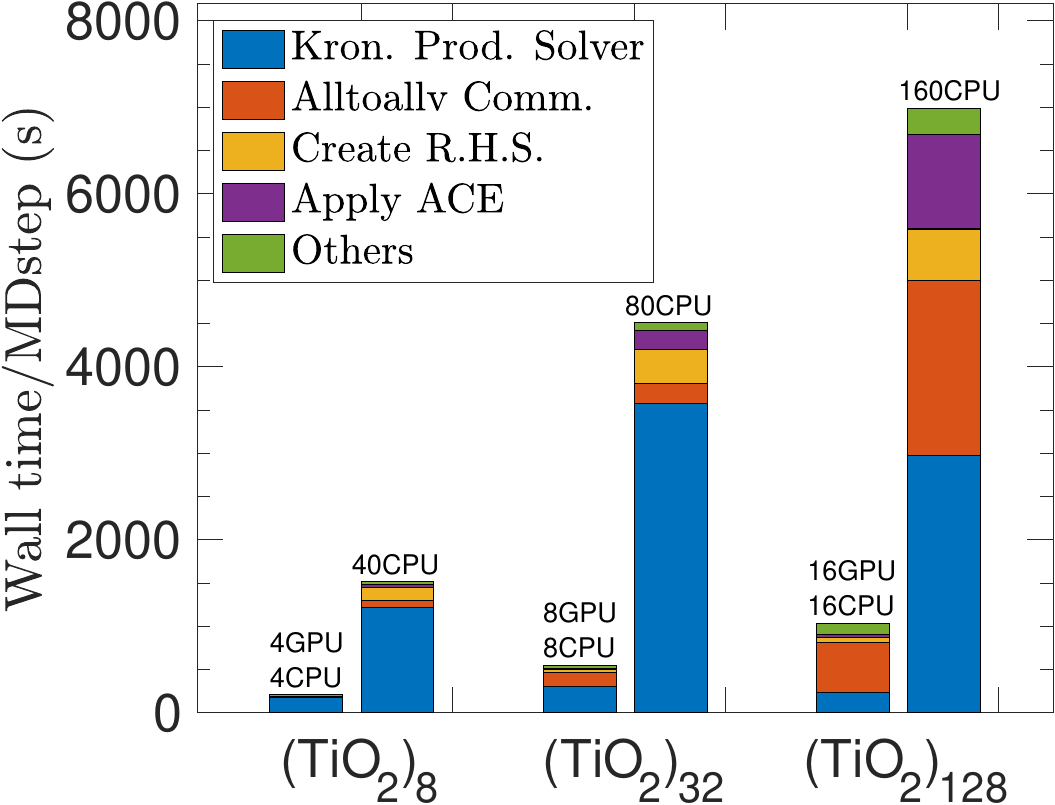} \label{Fig:TiO2_timesplit_min} }
\caption{\label{Fig:Timesplit_min} Breakdown of the timings for GPU-accelerated and CPU-only SPARC execution on the minimum number of computational nodes used in the strong scaling study (Fig.~\ref{Fig:Scaling}). The speedups in node hours for Mo$_{128}$, Mo$_{250}$, and Mo$_{432}$ in (Kron. Prod. Solver, \texttt{Alltoallv} Comm., Create R.H.S., Apply ACE, Others)  are (3.9x, 1.4x, 4.8x, 20.2x, 2.2x), (6.5x, 1.2x, 5.4x, 22.5x, 2.3x), and (5.9x, 2.2x, 3.1x, 30.4x, 2.7x), respectively. The corresponding numbers for (TiO$_2$)$_{8}$, (TiO$_2$)$_{32}$, and  (TiO$_2$)$_{128}$ are (6.9x, N/A, 25.9x, 11.3x, 1.4x), (11.6x, 1.4x, 13.8x, 18.4x, 2.4x), and (12.5x, 3.5x, 10.4x, 33.0x, 2.3x), respectively. For (TiO$_2$)$_{8}$ on 1 node, there is no domain parallelization and therefore no \texttt{Alltoallv} communication. The speedups are a factor of 10 larger in terms of core hours.}
\end{figure*}

\begin{figure*}[htbp!]
\centering
\subfloat[Bulk molybdenum]{\includegraphics[keepaspectratio=true,width=0.42\textwidth]{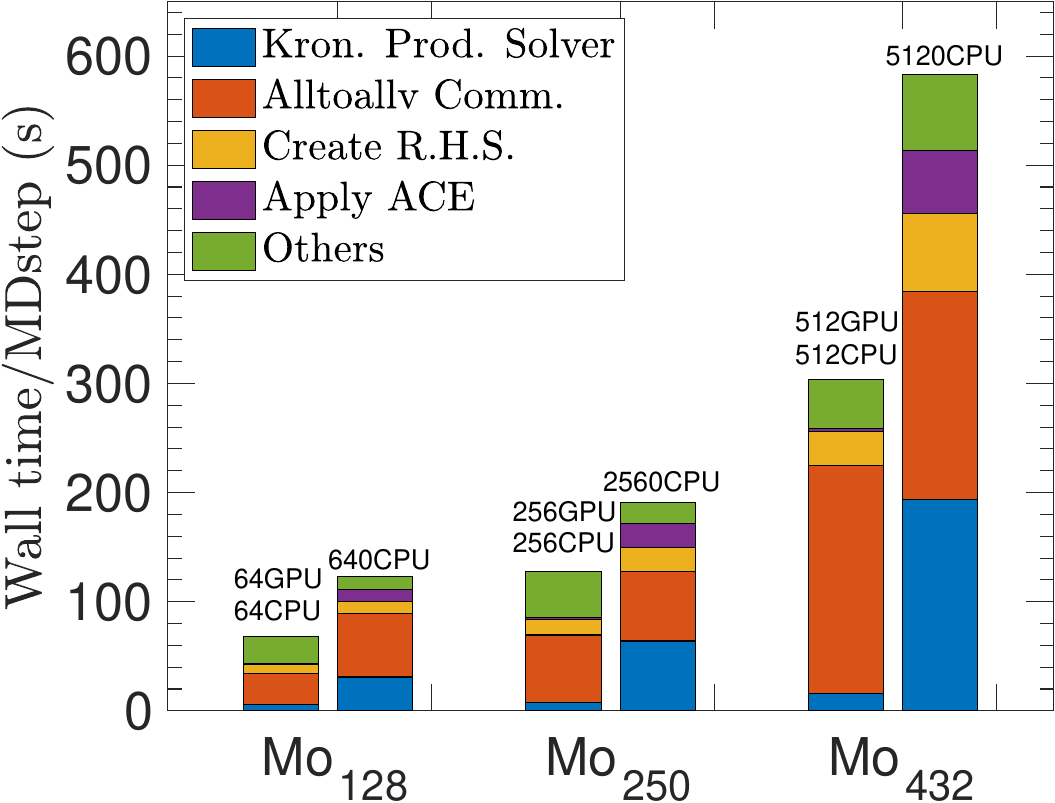} \label{Fig:Mo_timesplit_max} } \hspace{8mm}
\subfloat[(001) Titanium dioxide slab]{\includegraphics[keepaspectratio=true,width=0.42\textwidth]{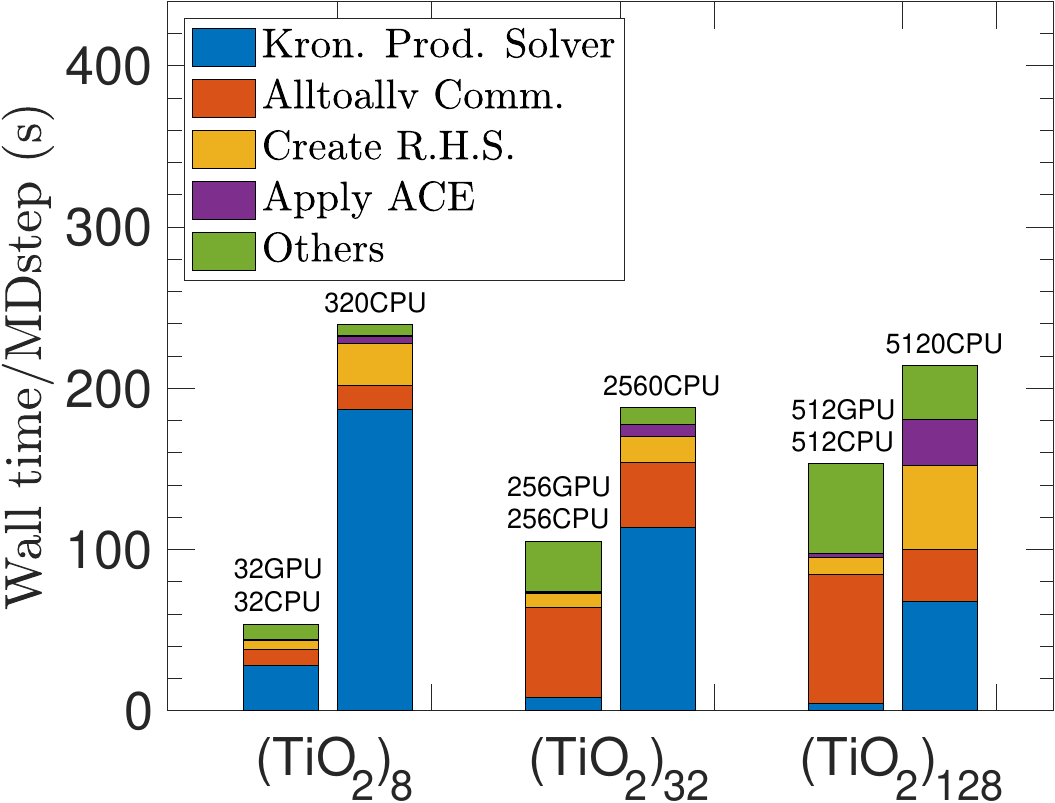} \label{Fig:TiO2_timesplit_max} }
\caption{\label{Fig:Timesplit_max} Breakdown of the timings for GPU-accelerated and CPU-only SPARC execution on the maximum number of computational nodes used in the strong scaling study (Fig.~\ref{Fig:Scaling}). The speedups in node hours for Mo$_{128}$, Mo$_{250}$, and Mo$_{432}$ in (Kron. Prod. Solver, \texttt{Alltoallv} Comm., Create R.H.S., Apply ACE, Others)  are (5.0x, 2.1x, 1.3x, 13.3x, 0.5x), (8.5x, 1.0x, 1.5x, 16.9x, 0.5x), and (12.1x, 0.9x, 2.3x, 21.9x, 1.6x), respectively. The corresponding numbers for (TiO$_2$)$_{8}$, (TiO$_2$)$_{32}$, and  (TiO$_2$)$_{128}$ are (6.7x, 1.5x, 4.4x, 7.3x, 0.8x), (13.9x, 0.7x, 1.8x, 7.4x, 0.3x), and (15.1x, 0.4x, 4.9x, 12.1x, 0.6x), respectively. The speedups are a factor of 10 larger in terms of core hours.}
\end{figure*}

To gain further insight into the performance of the GPU-accelerated version for hybrid functional calculations, we now analyze the timings associated with the key steps: Kronecker product solver, \texttt{Alltoallv} communication for the collection of the right hand side vectors, creation of the right hand side vectors that are suitable for the Kronecker product solver through Hadamard products and data reorganization, ACE operator application, and other parts including CheFSI routines. Since the time associated with the calculation of the atomic forces is negligible, it has been omitted from the analysis. In Figs.~\ref{Fig:Timesplit_min} and \ref{Fig:Timesplit_max}, we present the timing breakdown for GPU-accelerated and CPU-only executions on the minimum and maximum number of computational nodes used in the strong scaling study for each system (Fig.~\ref{Fig:Scaling}). We observe that the speedups for each step (other than the \texttt{Alltoallv} communication) are significantly larger on the minimum number of nodes compared to the maximum, due to the aforementioned processing capability of the GPUs. The performance of the custom \texttt{NCCL\_Alltoallv} routine relative to the \texttt{MPI\_Alltoallv} routine used in CPU-only execution follows the opposite trend, with speedups that range from 1.4x to 3.5x on the  maximum number of nodes, and from 0.4x to 2.1x on the minimum number of nodes. Notably, in terms of core hours the GPU implementation still achieves a speedup of over 4x  for the \texttt{Alltoallv} communication in all instances. The Kronecker product solver on the GPU consistently outperforms the solver on the CPU, with speedups ranging from 4x to 15x. The speedups in the creation of the right-hand side vectors range from 3.1x to 25.9x for the minimum number of computational nodes, and from 1.3x to 4.9x for the maximum number of nodes. TiO$_2$ systems typically exhibit greater speedups than Mo systems, primarily due to the larger number of finite-difference grid points in the TiO$_2$ systems, which allows them to take advantage of the aforementioned data processing ability of the GPUs. The speedups associated with the application of the ACE operator range from 11.3x to 30.4x on the minimum number of computational nodes, and from 7.3x to 21.9x on the maximum number of nodes. The CheFSI steps (without the application of the ACE operator) do not benefit as much from GPU acceleration as other steps, and are also slower compared to the results with local/semilocal functionals \cite{sharma2023gpu}.  This is because, with hybrid functionals, domain parallelization takes precedence over band parallelization, and GPU-GPU communication via \texttt{NCCL\_Neighbor\_alltoallv} becomes unavoidable in the discrete Laplacian kernel required for the evaluation of Hamiltonian-vector products, significantly reducing overall performance.

It is clear from the results presented above that the \texttt{Alltoallv} communication is the critical step limiting strong scaling efficiency on the \texttt{NVIDIA Volta V100} GPUs, and, consequently, the minimum time to solution that can be achieved. The time for this step is significantly reduced on using GPUs with larger bus bandwidth, e.g., the \texttt{Alltoallv} communication time is reduced by more than an order of magnitude on \texttt{NVIDIA H100} GPUs with \texttt{NVLink}, which offer an average bus bandwidth of 360 GB/s, whereas the \texttt{V100} node relies on Peripheral Component Interconnect Express (PCIe), with an average bandwidth of only 8 GB/s. Indeed, the \texttt{H100} delivers substantial performance gains across some other parts of the code, particularly in the Kronecker product solver. For example, consider two systems: (TiO$_2$)$_{32}$ and Mo$_{250}$, run on an HGX compute node with 8 \texttt{H100} GPUs. The overall speedup on the \texttt{H100} relative to the \texttt{V100} is 3.4x and 3.6x for (TiO$_2$)$_{32}$ and  Mo$_{250}$, respectively. Compared to CPU-execution, the overall speedup in node hours is 27.2x  and 14.4x for (TiO$_2$)$_{32}$ and Mo$_{250}$, respectively. The speedups in core hours are a factor of 8 larger. 

\section{Concluding Remarks} \label{Sec:Conclusions}
We have presented a GPU-accelerated version of the  SPARC real-space electronic structure code for hybrid functional calculations within generalized Kohn-Sham DFT. Specifically, we have developed a batch variant of the recently proposed Kronecker product linear solver for the solution of multiple linear systems simultaneously. We have then developed a modular, math kernel based implementation for hybrid functionals on NVIDIA architectures, offloading computationally intensive tasks to the GPUs while keeping the remaining workload on the CPUs. Considering bulk and slab examples, we have demonstrated that the GPU implementation can achieve up to 8x speedup in node-hours and 80x in core-hours relative to CPU-only execution, cutting the time to solution on \texttt{V100} GPUs to around 300 seconds for a metallic system with more than 6,000 electrons, while substantially reducing the computational resources needed for a given wall time. Opportunities for further reductions in wall time include  (i) developing an alternative to the \texttt{Alltoallv} communication scheme that involves the orbitals rather than right hand side vectors; (ii) utilizing the currently idle CPUs during GPU operation to handle some of the computations; and (iii) using more advanced  GPUs with larger bus bandwidth such as \texttt{NVIDIA H100}.

The modular and flexible design of the developed implementation enables straightforward extension to other GPU architectures, such as \texttt{AMD} and \texttt{Intel}, a direction that the authors are currently exploring. Other worthy subjects of research include extending the implementation to enable GPU acceleration for more advanced exchange–correlation functionals such as the random phase approximation correlation energy \cite{Shah2024}, which is significantly more computationally intensive than hybrid functionals; and GPU acceleration of the $\mathcal{O}(N)$ Spectral Quadrature (SQ) method \cite{suryanarayana2018sqdft}, enabling the study of systems with a million atoms \cite{gavini2022roadmap} and more, as next-generation parallel computing resources become available.

\begin{acknowledgments}
X.J. and P.S. gratefully acknowledge the support of the U.S. Department of Energy, Office of Science under grant DE-SC0023445. This work was performed in part under the auspices of the U.S. Department of Energy by Lawrence Livermore National Laboratory under Contract DE-AC52-07NA27344. J.E.P and A.S gratefully acknowledge support from the U.S. Department of Energy (DOE), National Nuclear Security Administration (NNSA): Advanced Simulation and Computing (ASC) Program at LLNL, and computational resources provided under the Multiprogrammatic and Institutional Computing programs at LLNL. This research was also supported by the supercomputing infrastructure provided by Partnership for an Advanced Computing Environment (PACE) through its Hive (U.S. National Science Foundation through grant MRI1828187) and Phoenix clusters at Georgia Institute of Technology, Atlanta, Georgia. X.J. acknowledges the help of Aaron Jezghani at PACE for the compilation of SPARC with \texttt{NVIDIA H100} GPUs.  
\end{acknowledgments}
\section*{Data Availability Statement}
The data that support the findings of this study are available within the article and from the corresponding author upon reasonable request.
\section*{Author Declarations}
The authors have no conflicts to disclose.
\section*{References}
\vspace{-5mm}
\bibliography{Manusript}

\end{document}